\documentclass[12pt]{iopart}
\usepackage{amsfonts}
\usepackage{graphicx}
\begin{document}
\title{Mathisson-Papapetrou-Dixon  equations in the Schwarzschild and Kerr backgrounds}
\author{Roman Plyatsko, Oleksandr Stefanyshyn, Mykola Fenyk}
\address{ Pidstryhach Institute of Applied Problems in Mechanics and
Mathematics,\\ Ukrainian National Academy of Sciences, 3-b Naukova
Str.,\\ Lviv, 79060, Ukraine}

\ead{plyatsko@lms.lviv.ua}

\begin{abstract}

A new representation, which does not contain the third-order
derivatives of the coordinates, of the exact
Mathisson-Papapetrou-Dixon equations, describing the motion of a
spinning test particle, is obtained under the assumption of the
Mathisson-Pirani condition in a Kerr background. For this purpose
the integrals of energy and angular momentum of the spinning
particle as well as a differential relationship following from the
Mathisson-Papapetrou-Dixon equations are used. The form of these
equations is adapted for their computer integration with the aim
to investigate the influence of the spin-curvature interaction on
the particle's behavior in the gravitational field without
restrictions on its velocity and spin orientation. Some numerical
examples for a Schwarzschild background are presented.

\end{abstract}


\maketitle
\section {Introduction}

In general relativity two main approaches have been developed for
the description of spinning particle behavior in a gravitational
field. Chronologically, the first one was initiated in 1929 when
the usual Dirac equation was generalized for curved space-time
[1]. The second, the pure classical (non-quantum) approach, was
proposed in 1937 [2]. Later it was shown that in a certain sense
the equations from [2] follow from the general relativistic Dirac
equation as a classical approximation [3].

The focus of this paper is on the equations of motion of the
classical spinning particle, which after [2] were obtained in [4]
and in many later papers by different methods. These equations can
be written as
\begin{equation}\label{1}
\frac D {ds} \left(mu^\lambda + u_\mu\frac {DS^{\lambda\mu}}
{ds}\right)= -\frac {1} {2} u^\pi S^{\rho\sigma}
R^{\lambda}_{~\pi\rho\sigma},
\end{equation}
\begin{equation}\label{2}
\frac {DS^{\mu\nu}} {ds} + u^\mu u_\sigma \frac {DS^{\nu\sigma}}
{ds} - u^\nu u_\sigma \frac {DS^{\mu\sigma}} {ds} = 0,
\end{equation}
where $u^\lambda\equiv dx^\lambda/ds$ is the particle's
4-velocity, $S^{\mu\nu}$ is the tensor of spin, $m$ and $D/ds$
are, respectively, the mass and the covariant derivative with
respect to the particle's proper time $s$;
$R^{\lambda}_{~\pi\rho\sigma}$ is the Riemann curvature tensor
(units $c=G=1$ are used); here and in the following, latin indices
run 1, 2, 3 and greek indices 1, 2, 3, 4; the signature of the
metric (--,--,--,+) is chosen.

Equations (1), (2) were generalized in [5] for the higher
multipoles of the test particles and now the set (1), (2) is known
as the Mathisson-Papapetrou-Dixon (MPD) equations. Multipolar
equations of motion for extended test bodies in general relativity
were considered in a recent paper [6] and in this context the
importance of the seminal work [2] was pointed out.

The first effects of the spin-gravity interaction following from
(1)--(2) were considered in [7] for the Schwarzschild field.
According to [7] and in many further publications (a list of them
is presented, for example, in [8, 9]) the influence of spin on the
particle's trajectory is negligible small for practical
registrations. However, in this sense much more realistic are the
effects connected with spin precession [10].

An interesting point has been elucidated in [11] concerning the
possibility of a static position of a spinning particle outside
the equatorial plane of the Kerr source of the gravitational
field, on its axis of rotation. In spite of the conclusion that
such a situation is not allowed by the MPD equations [11], this
question stimulated the investigations of possibilities of some
non-static (dynamical) effects connected with the particle's
motion relative to a Schwarzschild or Kerr mass outside the
equatorial plane [12]. Then it was shown that spinning particles
moving with relativistic velocity can significantly deviate from
geodesics [13, 14].

Some papers are devoted to the investigation of equilibrium
conditions of spinning test particles in the Schwarzschild-de
Sitter and Kerr-de Sitter space-times [15], as a development of
studying the stationary equilibrium positions of charged particles
in Reissner-Nordstrom and Kerr-Newman space-times [16].

While investigating the solutions of equations (1), (2), it is
necessary to add a supplementary condition  in order to choose an
appropriate trajectory of the particle's center of mass. Most
often conditions [2, 17]
\begin{equation}\label{3}
S^{\lambda\nu} u_\nu = 0
\end{equation}
or [5, 18]
\begin{equation}\label{4}
S^{\lambda\nu} P_\nu = 0
\end{equation}
are used, where
\begin{equation}\label{5}
P^\nu = mu^\nu + u_\lambda\frac {DS^{\nu\lambda}}{ds}
\end{equation}
is the 4-momentum. The condition for a spinning test particle
\begin{equation}\label{6} \frac{|S_0|}{mr}\equiv\varepsilon\ll 1
\end{equation}
must be taken into account as well [11], where  $|S_0|=const$ is
the absolute value of spin, $r$ is the characteristic length scale
of the background space-time (in particular, for the Kerr metric
$r$ is the radial coordinate), and $S_0$ is determined by the
relationship
\begin{equation}\label{7} S_0^2=\frac12
S_{\mu\nu}S^{\mu\nu}.
\end{equation}

In general, the solutions of equations (1), (2) under conditions
(3) and (4) are different. However, in the post-Newtonian
approximation these solutions coincide with high accuracy, just as
in some other cases [19, 20]. Therefore, instead of exact MPD
equations (1) their linear spin approximation
\begin{equation}\label{8}
m\frac D {ds} u^\lambda = -\frac {1} {2} u^\pi S^{\rho\sigma}
R^{\lambda}_{\,\,\,\pi\rho\sigma}
\end{equation}
is often considered. In this approximation condition (4) coincides
with (3) (by condition (3) $m$ in equations (1) is a constant
quantity).

According to [21] for a massless spinning particle, which moves
with the velocity of light, the appropriate condition is (3). The
question is which condition is adequate for the motions of a
spinning particle with the nonzero mass if its velocity is close
to the velocity of light? To answer this question it is necessary
to analyze the corresponding solutions of the exact MPD equations
(1), (2) both at condition (3) and (4).

The main purpose of this paper is to consider the exact MPD
equations under condition (3) in a Kerr metric. Due to the
symmetry of this metric equations (1), (2) have the constants of
motion: the particle's energy $E$ and the projection of its
angular momentum $J_z$ [22--25]. It is known that in the case of
the geodesic equations the analogous constants of motion were
effectively used for analyzing possible orbits of a spinless
particle in a Kerr space-time [26, 27]. Namely, by the constants
of energy and angular momentum the standard form of the geodesic
equations, which are the differential equations of the
second-order by the coordinates, can be reduced to the
differential equations of the first order. Naturally, it is
interesting to apply a similar procedure to the exact MPD
equations. However, in contrast to the geodesic equations, the
exact MPD equations at the condition (3) contain the third
derivatives of the coordinates [28, 29]. Therefore, the
application of this procedure to the exact MPD equations is
significant.

In this paper in order to obtain a full set of the MPD equations
at condition (3), without the third derivatives of the
coordinates, some differential relationship following from
equations (1), (2) is used. We present this relationship in
section 2 in general form, for any metric. Its concrete form in
the Kerr metric, together with the expressions for $E$ and $J_z$
is used in section 3 and the full set of the differential
equations for the dimensionless quantities connected with the
particle's coordinates, velocity and spin is described. The
explicit form of these equations are written in the Appendix. We
analyze the relationship between $u^\lambda$ and $P^\lambda$ at
condition (4) in section 4. Section 5 is devoted to some numerical
examples. We conclude in section 6.

It is important to note that the very condition (3) arose in a
natural fashion in the course of its derivation by different
methods [30--32]. Therefore, it is of importance to obtain a
representation of the exact MPD equations at this condition in the
Kerr metric convenient for their further computer integration.

We point out that the integrals of energy and angular momentum of
the MPD equations in a Kerr space-time were effectively used for
different purposes in [8, 22--25, 33--38] at condition (4).

 \vskip 3mm

 \section{ A relationship following from equations (1)--(3)}

In addition to the antisymmetric tensor $S^{\mu\nu}$ in many
papers the 4-vector of spin $s_\lambda$ is used as well, where by
definition
\begin{equation}\label{9}
s_\lambda=\frac12
\sqrt{-g}\varepsilon_{\lambda\mu\nu\sigma}S^{\nu\sigma}
\end{equation}
and $g$ is the determinant of the metric tensor,
$\varepsilon_{\lambda\mu\nu\sigma}$ is the Levi-Civita symbol. It
follows from (7), (9) that $s_\lambda s^\lambda=S_0^2$ and at
condition (3) we have $s_\lambda u^\lambda=0$ (other useful
relationships with $s_\lambda$ following from MPD equations at
different supplementary condition can be found, for example, in
[9]).

The set of equations (2) contains three independent differential
equations and in (3) we have three independent algebraic
relationships between $S^{\lambda\nu}$ and $u_\mu$. By (3) the
components $S^{i4}$ can be expressed through $S^{ki}$:
\begin{equation}\label{10}
S^{i4}=\frac{u_k}{u_4}S^{ki}.
\end{equation}
So, using (10) the components $S^{i4}$ can be eliminated both from
equations (2) and (1). That is, in further consideration one can
"forgot" about supplementary condition (3) and deal with the three
independent components $S^{ik}$. However, it is appear that more
convenient form of equations (1), (2) is not for $S^{ik}$ but for
another 3-component value  $S_i$ which is connected with $S^{ik}$
by the simple relationship
\begin{equation}\label{11} S_i=\frac{1}{2u_4}
\sqrt{-g}\varepsilon_{ikl}S^{kl},
\end{equation}
 where
$\varepsilon_{ikl}$ is the spatial Levi-Civita symbol. For
example, it is not difficult to check that three independent
equations of set (2) in terms of $S_i$ can be written as
\begin{equation}\label{12}
u_{4} \dot S_i  + 2(\dot u_{[4} u_{i]} - u^\pi u_\rho
\Gamma^\rho_{\pi[4} u_{i]})S_k u^k + 2S_n \Gamma^n _{\pi [4}
u_{i]} u^\pi =0,
\end{equation}
where a dot denotes usual differentiation with respect to the
proper time $s$, and square brackets denote antisymmetrization of
indices; $\Gamma^\rho_{\pi 4}$ are the Christoffel symbols.

The simple calculation shows that the 3-component value $S_i$ has
the 3-vector properties relative to the coordinate transformations
of the partial form $\hat x^i=\hat x^i(x^1, x^2, x^3), \quad \hat
x^4=x^4$ and in this special sense $S_i$ can be called as a
3-vector. (By the way, in the context of equations (1), (2)
firstly the 3-vector of spin was used in [7] with the notation
${\bf S}=(S^{23}, S^{31}, S^{12})$). By equations (9)--(11) the
relationship between $S_i$ and $s_\lambda$ is
\begin{equation}\label{13} S_i=-s_i+\frac{u_i}{u_4}s_4.
\end{equation}

Let us consider the first three equations of the subset (1) with
the indexes $\lambda=1, 2, 3$. Multiplying these equations by
$S_1,\quad S_2,\quad S_3$ correspondingly and taking into account
(9)--(11) we get
\begin{equation}\label{14}
mS_i\frac{Du^i}{ds}= -\frac12 u^\pi
S^{\rho\sigma}S_jR^j_{~\pi\rho\sigma}
\end{equation}
(here the covariant derivative $Du^i/ds$ remains 4D, i.e., is
determined according to the Christoffel connection of the
4-dimensional space-time).
 We stress that in contrast to the each equation from set
(1), which contain the third derivatives of the coordinates,
relationship (14) does not have these derivatives.

Relationship (14) is an analog of relationship (21) from [9] where
the spin 4-vector $s_\lambda$ is used.

\section{On set of exact MPD equations with constants
 of motion $E$, $J_z$ for the Kerr metric}
In the Boyer-Lindquist coordinates $x^1=r, \quad x^2=\theta, \quad
x^3=\varphi, \quad x^4=t$ the non-zero
 components of the Kerr metric tensor are
 \[
g_{11}=-\frac{\rho^2}{\Delta}, \quad g_{22}=-\rho^2,
\]
\[
g_{33}=-\left(r^2+a^2+\frac{2Mra^2}{\rho^2}
\sin^2\theta\right)\sin^2\theta,
\]
\begin{equation}\label{15}
 g_{34}=\frac{2Mra}{\rho^2}\sin^2\theta, \quad
g_{44}=1-\frac{2Mr}{\rho^2},
\end{equation}
where
\[
 \rho^2=r^2+a^2\cos^2\theta, \quad \Delta=r^2-2Mr+a^2, \quad
0\le\theta\le\pi.
\]

In this coordinates the constant values of the particle's energy
$E$ and the projection of its angular momentum $J_z$ can be
written as [22--25]
\begin{equation}\label{16}
E=P_4-\frac{1}{2}g_{4\mu,\nu}S^{\mu\nu},
\end{equation}
 \begin{equation}\label{17}
J_z=-P_3+\frac{1}{2}g_{3\mu,\nu}S^{\mu\nu}.
\end{equation}

It is convenient to use the dimensionless quantities $y_i$
connected with the  particle's coordinates by definition
\begin{equation}\label{18}
\quad y_1=\frac{r}{M},\quad y_2=\theta,\quad y_3=\varphi, \quad
y_4=\frac{t}{M},
\end{equation}
as well as the quantities connected with its 4-velocity
\begin{equation}\label{19}
y_5=u^1,\quad y_6=Mu^2,\quad y_7=Mu^3,\quad y_8=u^4
\end{equation}
and the spin components [14]
\begin{equation}\label{20}
    y_9=\frac{S_1}{mM},\quad y_{10}=\frac{S_2}{mM^2},\quad
    y_{11}=\frac{S_3}{mM^2}.
\end{equation}
(We underline that in the following, in sections 4, 5 and in the
Appendix, the notation $y_i^k$ means that the quantity number $i$
from the set of eleven quantities (18)--(20) is to the $k$-th
power.)
 In addition, we introduce the dimensionless quantities
connected with the particle's proper time $s$ and the constants of
motion $E$ , $J_z$:
\begin{equation}\label{21}
    x=\frac{s}{M}, \quad \hat E=\frac{E}{m},\quad
    \hat J=\frac{J_z}{mM}.
\end{equation}
Quantities (18), (19) satisfy the four simple equations
\begin{equation}\label{22}
\dot y_1=y_5, \quad \dot y_2=y_6, \quad \dot y_3=y_7, \quad \dot
y_4=y_8,
\end{equation}
here and in the following a dot denotes the usual derivative with
respect to $x$.

Now we point out other seven nontrivial first-order differential
equations for the 11 functions $y_i$. Namely, the first of them
follows directly from equation (14). The second is a result of the
covariant differentiation of the normalization condition $u_\nu
u^\nu=1$, that is
\begin{equation}\label{23}
u_\nu\frac{Du^\nu}{ds}=0.
\end{equation}
The third and fourth equations follow from (16) and (17)
correspondingly if condition (3) is taken into account. Finally,
the last three equations for $y_i$ follow directly from (12). This
set of the seven equations is presented in the Appendix. Equations
(A.3)--(A.9) together with the four equations from (22) are the
full set of the exact MPD equations which describe most general
motions of a spinning particle in the Kerr gravitational field
without any restrictions on its velocity and spin orientation. We
stress that the two equations following from (16) and (17) (see
(A.5), (A.6)) contain the quantities $\hat E$ and $\hat J$ as the
parameters proportional to the particle's energy and angular
momentum according to notation (21).

Now we recall some features of the solutions of the exact MPD
equations under condition (3). It is known that in the Minkowski
space-time equations (1)--(3) have, in addition to usual solutions
describing the straight worldlines, a set of solutions describing
the oscillatory (helical) worldlines [28, 29]. The physical
interpretation of these superfluous solutions was proposed by C.
M{\"o}ller [30]. He pointed out that in relativity the position of
the center of mass of a rotating body depends on the frame of
reference, and condition (3) is common for the so-called proper
and non-proper centers of mass [39]. The usual solution describe
the motion of the proper center of mass and the helical solutions
describe the motions of the set of the non-proper centers of mass.
Naturally, in general relativity, when the gravitational field is
present, the exact MPD equations (1)--(3) have some superfluous
solutions as well. Just to avoid these solutions, instead of (3)
condition (4) was used in many papers. In contrast to (3),
condition (4) picks out the unique worldline of a spinning
particle in the gravitational field. However, the question arises:
is this worldline close, in the certain sense, to the usual
(non-helical) worldline of equations (1), (2) under condition (3)?
It is simple to answer this question if the linear spin
approximation is valid, because in this case condition (4)
practically coincides with (3). Whereas another situation cannot
be excluded {\it a priori} for the high particle's velocity.

Concerning equations (22), (A.3)--(A.9) we stress that by choosing
different values of $\hat E$ and $\hat J$ for the fixed initial
values of $y_i$ one can describe the motions of different centers
of mass. Among the set of the pairs $\hat E$ and $\hat J$ there is
the single pair corresponding to the proper center of mass. The
possible approaches for finding this pair is a separate subject.
One of them was proposed in [40] where a method of separation of
some non-oscillatory solutions of the exact MPD equations in the
Schwarzschild field was considered. In the next section we shall
analyze the possibility of using the expressions for $E$ and $J_z$
from (16), (17) at condition (4) for the same purpose. Note that
under condition (4) the concrete values of $E$ and $J_z$ are fully
determined by the initial values of the particle coordinates and
velocity (or momentum), without its acceleration, in contrast to
the case with condition (3).

\section{Values $\hat E$ and $\hat J$ according to condition (4)}

Let us check the supposition that the single solutions of
equations (1), (2) at the supplementary condition (4),
corresponding to the fixed initial values of the coordinates,
velocity and spin, is close to those solutions of equations (1),
(2) at the condition (3) which describe the motion of the proper
center of mass with the same initial values. As we pointed out in
section 1, this assumption is justified for the velocities which
are not close to the velocity of light. Here we shall consider the
situation for any velocity.

Let us write the main relationships following from the MPD
equations at condition (4) [22--25]. The mass of a spinning
particle $m$ is defined as
\begin{equation}\label{24}
m=\sqrt{P_\lambda P^\lambda}
\end{equation}
and $m$ is the integral of motion, that is $dm/ds=0$. The quantity
$V^\lambda$ is the normalized momentum, where by definition
\begin{equation}\label{25}
V^\lambda=\frac{P^\lambda}{m}.
\end{equation}
Sometimes $V^\lambda$ is called the "dynamical 4-velocity",
whereas the quantity $u^\lambda$ from (1)--(3) is the "kinematical
4-velocity" [8]. As the normalized quantities $u^\lambda$ and
$V^\lambda$ satisfy the relationships
\begin{equation}\label{26}
u_\lambda u^\lambda=1,\quad V_\lambda V^\lambda=1.
\end{equation}
There is the important relationship between $u^\lambda$ and
$V^\lambda$ [22--24]:
\begin{equation}\label{27}
    u^{\lambda}=N\left[V^\lambda+\frac{1}{2m^2\Delta}
    S^{\lambda\nu}V^{\pi}R_{\nu\pi\rho\sigma}S^{\rho\sigma}\right],
\end{equation}
where
\begin{equation}\label{28}
\Delta=1+\frac{1}{4m^2}R_{\lambda\pi\rho\sigma}S^{\lambda\pi}S^{\rho\sigma},
\end{equation}

Now our aim is to consider the explicit form of expression (27)
for the concrete case of the Schwarzschild metric, for the
particle motion in the plane $\theta=\pi/2$ when spin is
orthogonal to this plane (we use the standard Schwarzschild
coordinates $x^1=r, \quad x^2=\theta, \quad x^3=\varphi, \quad
x^4=t$). Then we have
\begin{equation}\label{29}
u^2=0,\quad u^1\ne 0, \quad u^3\ne 0, \quad u^4\ne 0,
\end{equation}
\begin{equation}\label{30}
S^{12}=0,\quad S^{23}=0,\quad S^{13}\ne 0.
\end{equation}
In addition to (30) by condition (4) we write
\begin{equation}\label{31}
 S^{14}=-\frac{P_3}{P_4}S^{13},\quad S^{24}=0, \quad S^{34}=\frac{P_1}{P_4}S^{13}.
\end{equation}
Using (7), (29)--(31) and the corresponding expressions for the
Riemann tensor in the Schwarzschild metric, from (27) we obtain
\[
u^1=NV^1\left(1+\frac{3M}{r^3}V_3
V^3\frac{S_0^2}{m^2\Delta}\right), \quad u^2=V^2=0,
\]
\[
u^3=NV^3\left[1+\frac{3M}{r^3}(V_3
V^3-1)\frac{S_0^2}{m^2\Delta}\right],
\]
\begin{equation}\label{32}
u^4=NV^4\left(1+\frac{3M}{r^3}V_3
V^3\frac{S_0^2}{m^2\Delta}\right),
\end{equation}
where $M$ is the mass of a Schwarzschild source. According to (28)
we write the expression for $\Delta$ as
\begin{equation}\label{33}
\Delta=1+\frac{S_0^2 M}{m^2 r^3}(1-3V_3 V^3)
\end{equation}
(the quantity $M$ in (32), (33) is the mass of a Schwarzschild
source). Inserting (33) into (32) we get
\[
u^1=\frac{NV^1}{\Delta}\left(1+\frac{S_0^2 M}{m^2 r^3}\right),
\]
\[u^3=\frac{NV^3}{\Delta}\left(1-2\frac{S_0^2 M}{m^2 r^3}\right),
\]
\begin{equation}\label{34}
u^4=\frac{NV^4}{\Delta}\left(1+\frac{S_0^2 M}{m^2 r^3}\right).
\end{equation}
As in (6), we note
\begin{equation}\label{35}
\varepsilon=\frac{|S_0|}{mr},
\end{equation}
where according to the condition for a test particle it is
necessary $\varepsilon\ll 1.$ However, in our calculations we
shall keep all terms with $\varepsilon.$

 The explicit expressions for $N$ we obtain directly from the condition
$u_\lambda u^\lambda=1$ in the form
\[
N=\Delta\left[\left(1+\varepsilon^2\frac{M}{r}\right)^2-3V_3
V^3\varepsilon^2\frac{M}{r}\times\right.
\]
\begin{equation}\label{36}
\left.\times\left(2-\varepsilon^2\frac{M}{r}\right)\right]^{-1/2}.
\end{equation}
Inserting (36) into (34) we obtain the expression for the
components $V^\lambda$ through $u^\lambda$ ($V^2=u^2=0$):
\[
V^1=u^1R\left(1-2\varepsilon^2\frac{M}{r}\right),
\]
\[
V^3=u^3R\left(1+\varepsilon^2\frac{M}{r}\right),
\]
\begin{equation}\label{37}
V^4=u^4R\left(1-2\varepsilon^2\frac{M}{r}\right),
\end{equation}
where
\begin{equation}\label{38}
R=\left[\left(1-2\varepsilon^2\frac{M}{r}\right)^2-3(u^3)^2\varepsilon^2
Mr\left(2-\varepsilon^2\frac{M}{r}\right)\right]^{-1/2}.
\end{equation}

The main feature of relationships (37), (38) is that for the high
tangential velocity of a spinning particle the values $V^1, V^3,
V^4$ become imaginary. Indeed, if
\begin{equation}\label{39}
|u^3|>\frac{1}{\varepsilon\sqrt{6Mr}},
\end{equation}
in (38) we have the square root of the negative value. (As writing
(39) we neglect the small terms of order $\varepsilon^2$; all
equations in this section before (39) and after (40) are strict in
$\varepsilon.$) Using the notation for the particle's tangential
velocity $u_{tang}\equiv ru^3$ by (39) we write
\begin{equation}\label{40}
|u_{tang}|>\frac{\sqrt{r}}{\varepsilon\sqrt{6M}}.
\end{equation}
According to estimates similar to those which are presented in
[14] if $r$ is not much greater than $M$, the velocity value of
the right-hand side of equation (40) corresponds to the particle's
highly relativistic Lorentz $\gamma$-factor of order
$1/\varepsilon$.

Probably, this fact that according to (25), (37)--(40) the
expressions for the components of 4-momentum $P^\lambda$ become
imaginary (if $m$ in (24) is real) is an evidence that condition
(4) cannot be used for the particle's velocity which is very close
to the velocity of light. However, this point needs some
additional consideration. In any case, relationships (37)--(40)
are of importance for authors which investigate solutions of the
MPD equations at condition (4). We stress that many papers were
devoted to study the planar or circular motions of spinning
particles in the Schwarzschild or Kerr space-time at different
supplementary conditions [7--14, 22--25, 33--35, 40, 41 ].
Equations (37)--(40) elucidate the new specific features which
arise for the highly relativistic motions.

Another aspect of the connection between the spinning particle's
momentum and velocity at condition (4) was considered in [22, 23].
It is pointed out in [22] that there exist a critical distance of
minimum approach of a particle to the Kerr source where its
velocity becomes space-like with the time-like momentum. Similarly
according to conclusion from [23] the velocity of the spinning
particle become space-like for sufficiently large gravitational
fields and/or spins. In contrast to [22, 23] expressions
(37)--(40) describe the situation when the velocity is time-like
but the components of momentum become imaginary.

The new result of this section as compare to [22, 23] consists in
the conclusion that according to (37)--(40) only tangential
component of velocity is important in this case, not the radial
one, although the orbit is not necessarily circular.

It is interesting to check the possibility of using the values $E$
and $J_z$ as calculated by (37) for computing spinning particle
motions by the equations which are described in the previous
section if the particle's tangential velocity is much less than
the critical value from the right-hand side of (40). At condition
(4) the constants $E$ and $J_z$ for the equatorial motions in the
Schwarzschild field can be written as
\begin{equation}\label{41}
E=P_4+\frac{1}{2}g_{44,1}S^{14}=m
V_4-\frac{1}{2}g_{44,1}\frac{V_3}{V_4}S^{13},
\end{equation}
 \begin{equation}\label{42}
J_z=-P_3-\frac{1}{2}g_{33,1}S^{13}=-m
V_3-\frac{1}{2}g_{33,1}S^{13}.
\end{equation}
Using the dimensionless quantities $y_i$ as defined in (18), (19),
relationship (37) and the simple expression for $S^{13}$ through
$S_0$ from (41), (42) we obtain
\begin{equation}\label{43}
\hat
E=R\left[y_8\left(1-\frac{2}{y_1}\right)\left(1-\frac{2\varepsilon^2}{y_1}\right)+
\varepsilon y_7\left(1+\frac{\varepsilon^2}{y_1}\right)\right],
\end{equation}
\begin{equation}\label{44}
\hat J=R\left[y_1^2 y_7 \left(1+\frac{\varepsilon^2}{y_1}\right) +
\varepsilon y_1 y_8
\left(1-\frac{2}{y_1}\right)\left(1-\frac{2\varepsilon^2}{y_1}\right)\right],
\end{equation}
where according to (21) we note $\hat E=E/m$, $\hat J=J_z/mM$.
Then for the fixed initial values of the quantities $y_1, y_7,
y_8$ in (43), (44) we have the concrete values of $\hat E$ and
$\hat J$ which can be used for numerical integration of the exact
MPD equations. Some examples we shall consider in the next
section.

\section{Numerical examples}

Let us consider some solutions of the exact MPD equations (22),
(A.3)--(A.9) for the equatorial particle's motion in the
Schwarzschild field with the values $\hat E$, $\hat J$ from (43),
(44). We are interested in the highly relativistic motions when
$u_{tang}^2\gg 1$ and, at the same time, $|u_{tang}|$ is much less
than right-hand side of (40). All figures 1--8 correspond to the
situation when the initial values of the particle coordinates and
velocity are given as
\begin{equation}\label{45}
r(0)=2.5M, \varphi (0)=0, u^1(0)=-1, Mu^3(0)=100
\end{equation}
(for the equatorial motions $\theta=\pi/2$ identically, i.e.,
$u^2=0$; $u^4(0)$ is determined from the condition $u_\nu
u^\nu=1$). The small value $S_0/Mm$ is equal to $10^{-6}$ for
figures 1--3 and $6\times 10^{-5}$ for figures 4--8. According to
(35) for the spin 3-vector we have $S_1=0$, $S_3=0$ and it follows
from (12) that $S_2=rS_0$. For comparison, we present the
corresponding solutions of the geodesic equations with the same
initial values of the coordinates and velocity.

Figures 1 and 4 show the dependence  $r$ vs. $s$, with the
difference that in figure 1 the graph for a spinning particle
practically coincides with the corresponding geodesic graph,
whereas according to figure 4 the graph for a spinning particle
reveals the oscillatory features. The similar property takes place
for the dependence  $\varphi$ vs. $s$ according to figure 5 (the
simple linear dependence $\varphi$ vs. $s$  for a spinning
particle with $S_0/Mm=10^{-6}$ is not presented here). Figures 2,
3, 6, 7 illustrate the clear oscillatory regime for the radial and
angular particle's velocity. We point out that the amplitude of
the oscillation increases with the value of spin, whereas the
frequency of this oscillation decreases. At the same time,
according to figure 8 the dependence $r$ vs. the coordinate time
$t$ does  not reveal the oscillatory features. It is not strange
because $dt/ds$ oscillates similarly to $dr/ds$ and $d\varphi/ds$
(the corresponding graph for $dt/ds$ is not presented here for
brevity) in such a manner that $dr/dt$ is not oscillatory.

It is known that non-proper centers of mass of a spinning particle
oscillate around the proper center of mass [28, 29, 39].
Therefore, one can consider the middle lines of the corresponding
oscillatory lines in figures 2--7 as such that present the motions
of the proper center of mass. So, when $|u_{tang}|$ is much less
than right-hand side of (45), the figures above show that
equations (22), (A.3)--(A.9) with the relationships (43), (44) can
be used to describe the spinning particle motions in the
Schwarzschild space-time in some approximation.

Because of the specific oscillatory regime the corresponding
graphs in figures 1--8 are calculated on the limited proper time
interval. To illustrate some longer trajectories we shall use the
expressions for $\hat E$ and $\hat J$ which follows from
relationships (34)--(43) of paper [40]. We recall that in [40] an
approach for separation of some non-oscillatory solutions of the
exact MPD equations at condition (3) in the equatorial plane of
the Schwarzschild space-time was developed (these solutions
describe the orbits closer to circular with $r<3M$). So, according
to our notation $y_1=r/M$, $\hat E=E/m$, $\hat J=J_z/mM$, and
$\varepsilon_0=S_0/Mm$, now we rewrite the corresponding
expressions which are presented in equations (34)--(43) of [40] as
\begin{equation}\label{46}
Mu^3=-\frac{1}{\sqrt{\varepsilon_0 y_1}}
\left(1-\frac{2}{y_1}\right)^{1/4}\left|1-\frac{3}{y_1}\right|^{-1/2},
\end{equation}
\begin{equation}\label{47}
\hat E=\frac{\sqrt{\varepsilon_0}}{\sqrt{y_1}}
\left(1-\frac{2}{y_1}\right)^{1/4}\left|1-\frac{3}{y_1}\right|^{-3/2}
\left(1-\frac{3}{y_1}+\frac{3}{y_1^2}\right),
\end{equation}
\begin{equation}\label{48}
\hat J= \sqrt{\varepsilon_0
y_1}\left(1-\frac{2}{y_1}\right)^{-1/4}\left|1-\frac{3}{y_1}\right|^{-3/2}
\left(1-\frac{9}{y_1}+\frac{15}{y_1^2}\right).
\end{equation}

Figures 9--11 correspond to the situation when the initial values
of the particle coordinates and velocity are given as
\begin{equation}\label{49}
r(0)=2.8M, \varphi (0)=0, u^1(0)=-0.5
\end{equation}
($u^2=0$ identically). In addition we put $\varepsilon_0=10^{-6}$,
then according to (46) we have $Mu^3\approx -1635$. For comparison
in figures 9--11 the corresponding graphs for the geodesic motion
are presented. Figures 9--11 illustrate the significant space
separation of the trajectories of spinning and spinless particles:
for the proper time $s\approx 1.3\times 10^{-3}M$ (it corresponds
to the coordinate time $t\approx 25M$) the spinless particle falls
on the horizon surface after less than 0.5 revolution by the angle
$\varphi$ about the Schwarzschild mass, whereas the position of
the spinning particle by the radial coordinate is close to the
initial value $r=2.8M$  for one revolution.

We point out that figures 9--11 are similar to same figures from
our paper [14]. However, the essential difference must be
stressed: all figures in [14] are presented for the linear spin
approximation of the MPD equations, whereas here we deal with the
exact equations.

\begin{figure}[!]
\centering
\includegraphics[width=6cm]{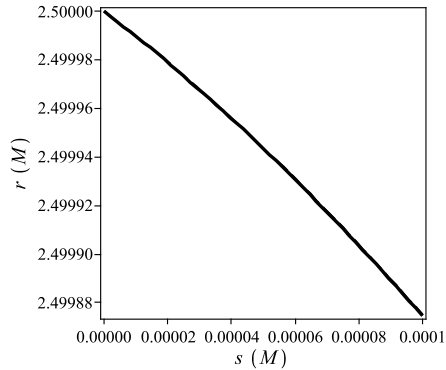}
\caption{\label{1} Radial coordinate vs. proper time at
$S_0/Mm=10^{-6}$. The line for the spinning particle practically
coincides with the geodesic line at the same initial values (45).
}
\end{figure}

\begin{figure}[!]
\centering
\includegraphics[width=6cm]{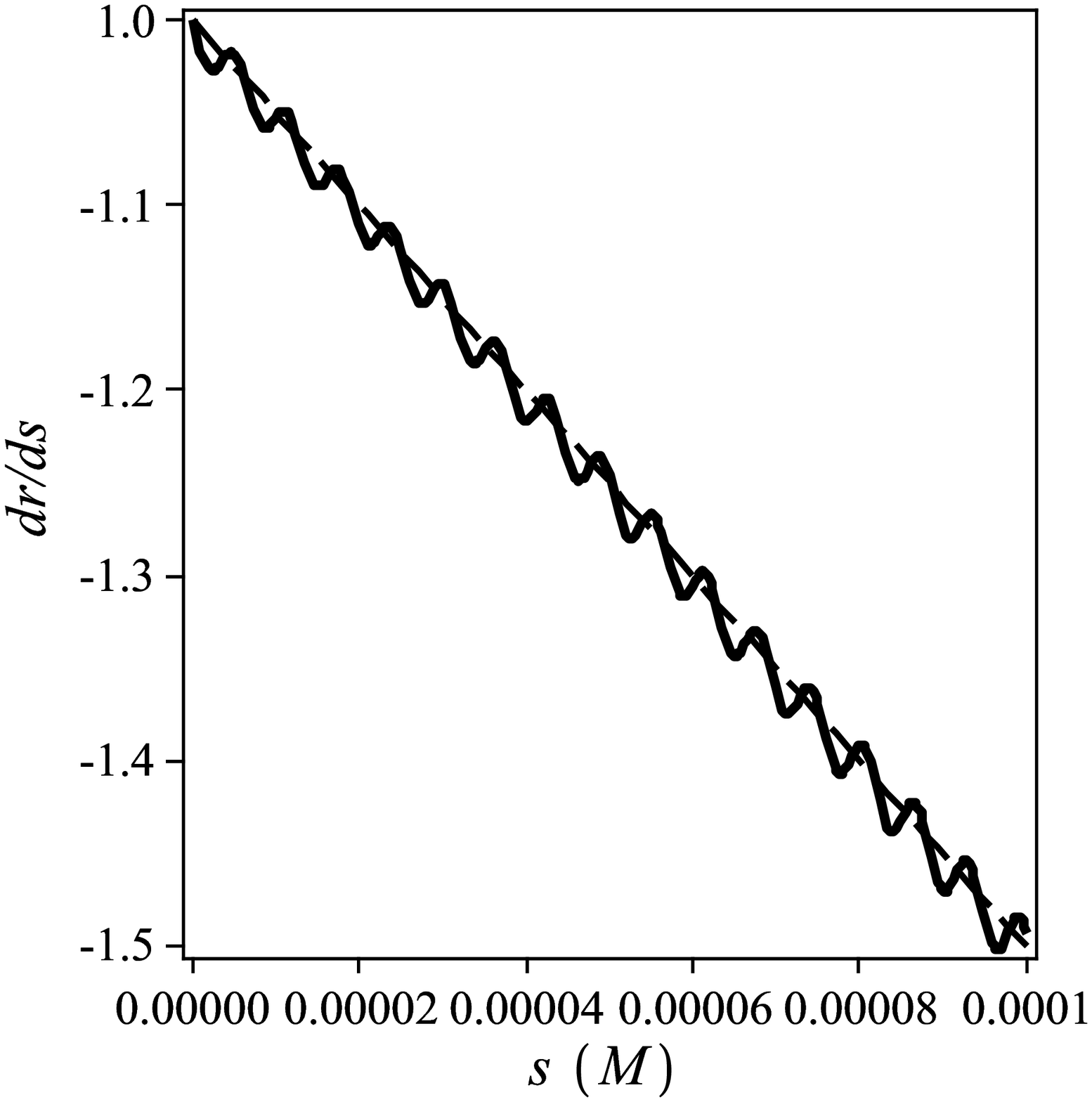}
\caption{\label{1} Radial velocity vs. proper time for the
spinning particle with $S_0/Mm=10^{-6}$ (solid line) and for the
geodesic motion (dashed line).}
\end{figure}

\begin{figure}[!]
\centering
\includegraphics[width=6.5cm]{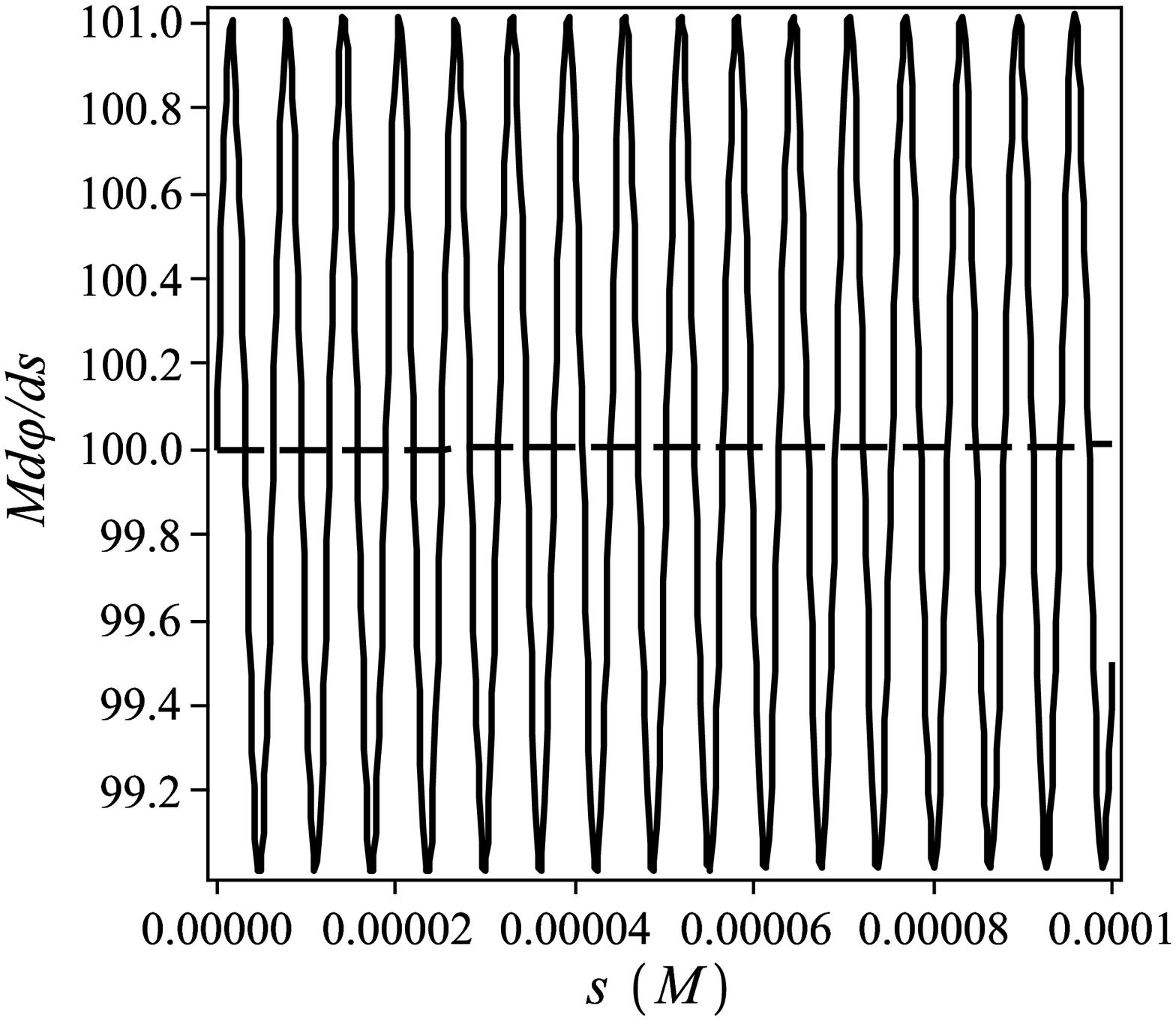}
\caption{\label{1} Angular velocity vs. proper time for the
spinning particle with $S_0/Mm=10^{-6}$ (solid line) and for the
geodesic motion (dashed line).}
\end{figure}

\begin{figure}[!]
\centering
\includegraphics[width=6cm]{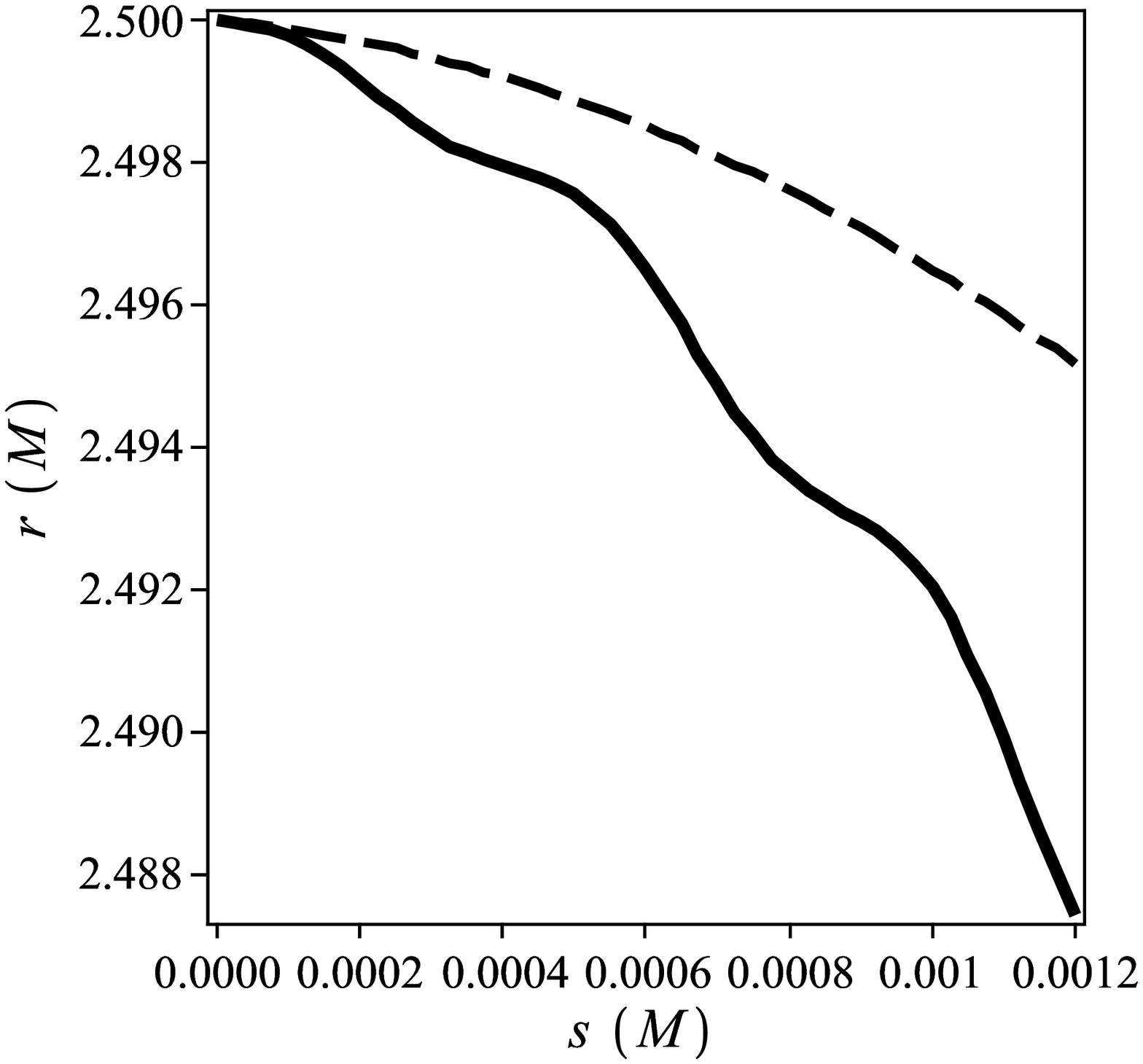}
\caption{\label{1} Radial coordinate vs. proper time for the
spinning particle with $S_0/Mm=6\times 10^{-5}$ (solid line) and
for the geodesic motion (dashed line).}
\end{figure}

\begin{figure}[!]
\centering
\includegraphics[width=6cm]{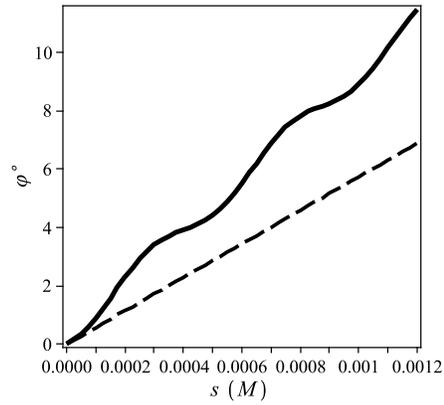}
\caption{\label{1} Angle $\varphi$ vs. proper time for the
spinning particle with $S_0/Mm=6\times 10^{-5}$ (solid line) and
for the geodesic motion (dashed line).}
\end{figure}

\begin{figure}[!]
\centering
\includegraphics[width=6.3cm]{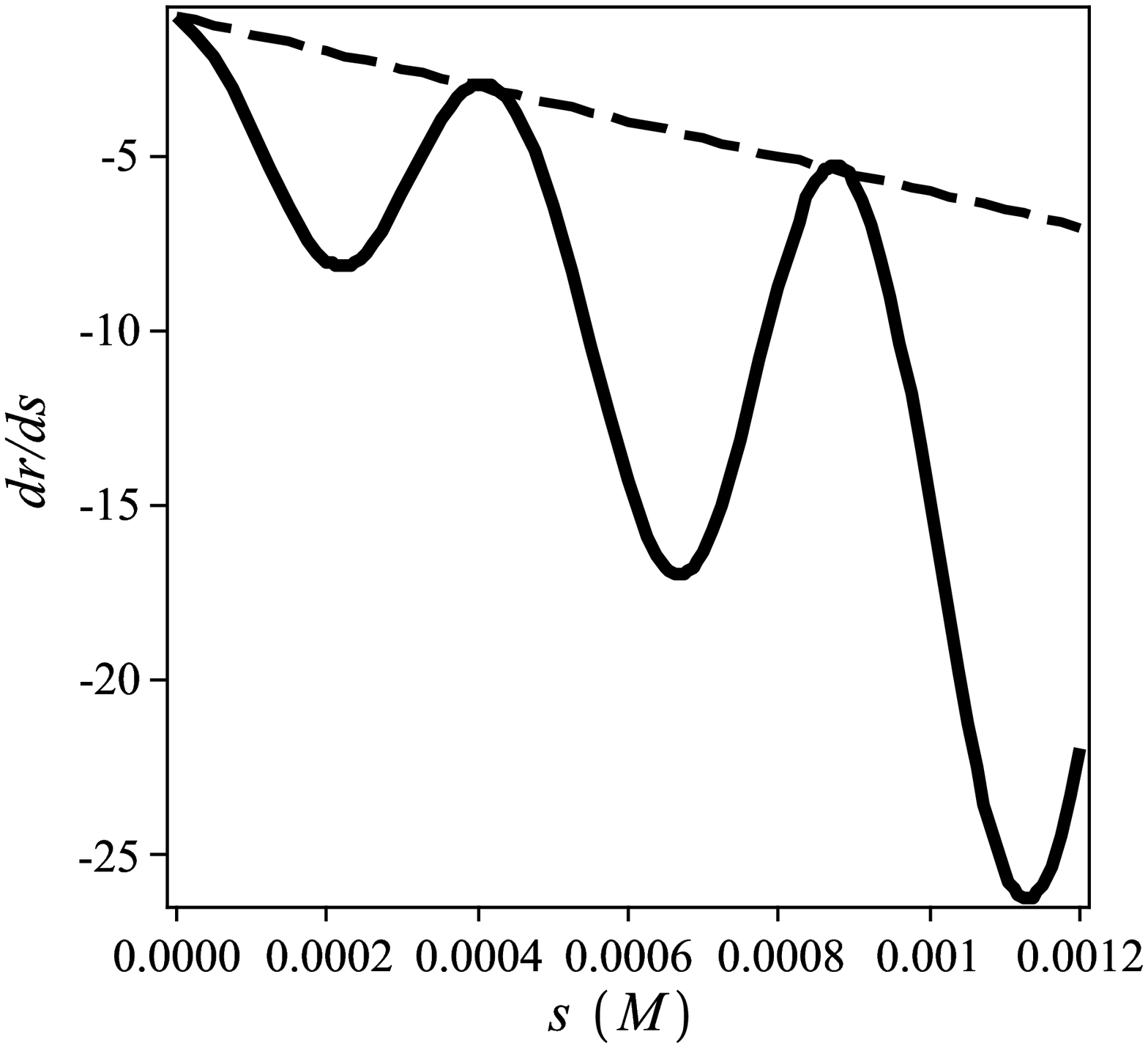}
\caption{\label{1} Radial velocity vs. proper time for the
spinning particle with $S_0/Mm=6\times 10^{-5}$ (solid line) and
for the geodesic motion (dashed line).}
\end{figure}

\begin{figure}[!]
\centering
\includegraphics[width=6.5cm]{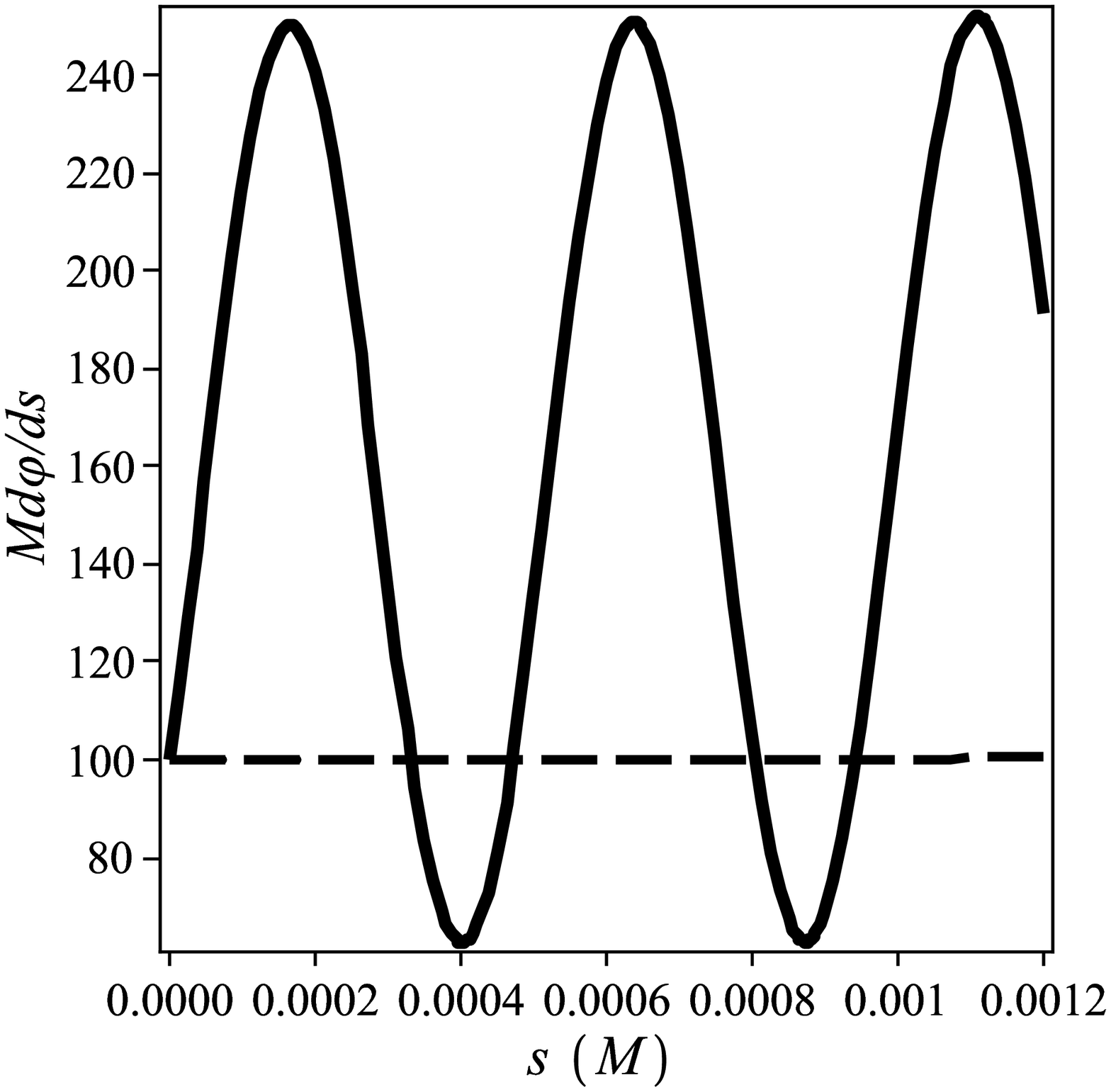}
\caption{\label{1} Angular velocity vs. proper time for the
spinning particle with $S_0/Mm=6\times 10^{-5}$ (solid line) and
for the geodesic motion (dashed line).}
\end{figure}

\begin{figure}[!]
\centering
\includegraphics[width=6.5cm]{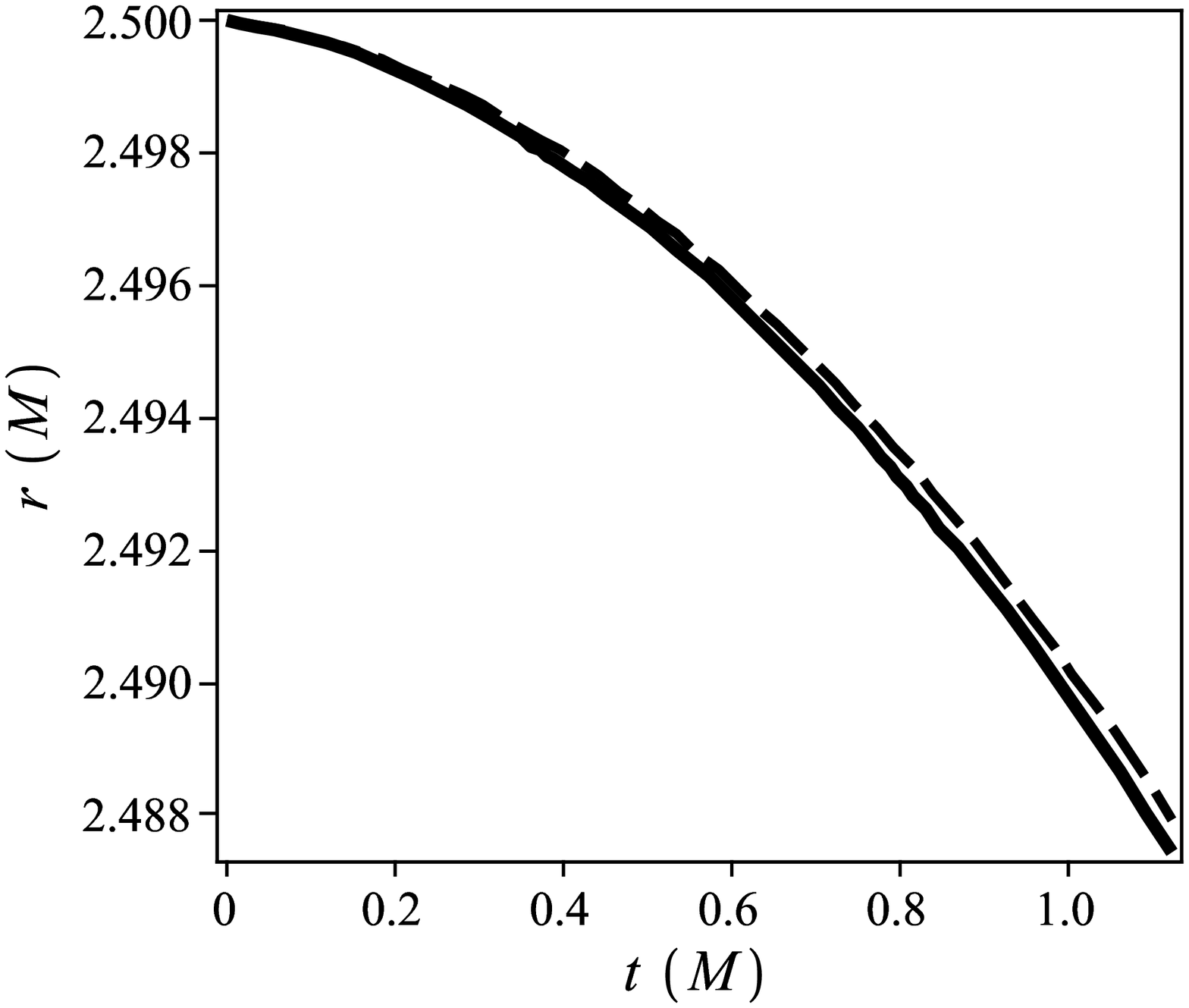}
\caption{\label{1} Radial coordinate vs. coordinate time for the
spinning particle with $S_0/Mm=6\times 10^{-5}$ (solid line) and
for the geodesic motion (dashed line).}
\end{figure}

\begin{figure}[!]
\centering
\includegraphics[width=6.5cm]{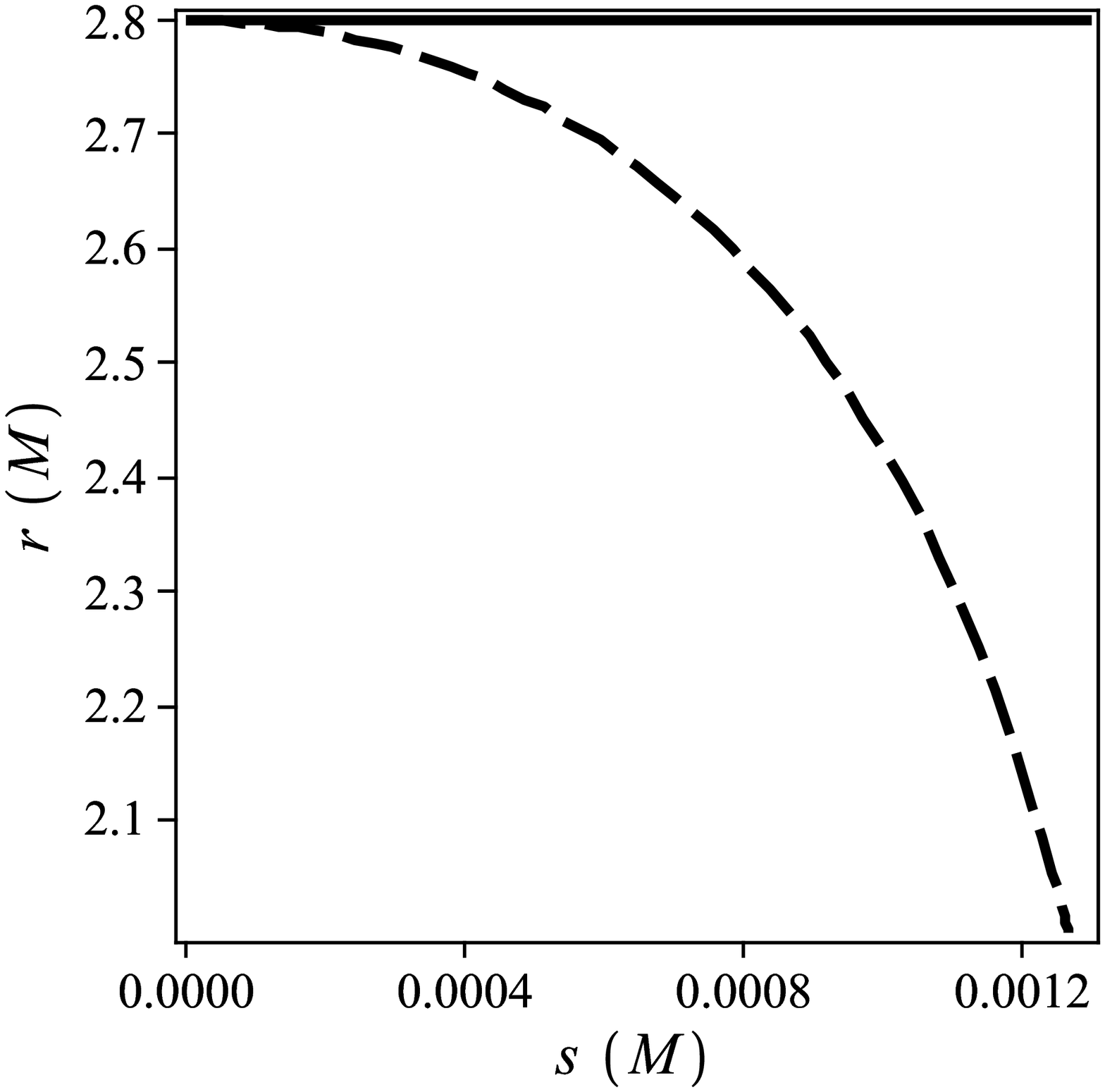}
\caption{\label{1} Radial coordinate vs. proper time for the
spinning particle with $S_0/Mm= 10^{-6}$ (solid line) and for the
geodesic motion with the same initial values of the coordinates
and velocity (dashed line).}
\end{figure}

\begin{figure}[!]
\centering
\includegraphics[width=6.5cm]{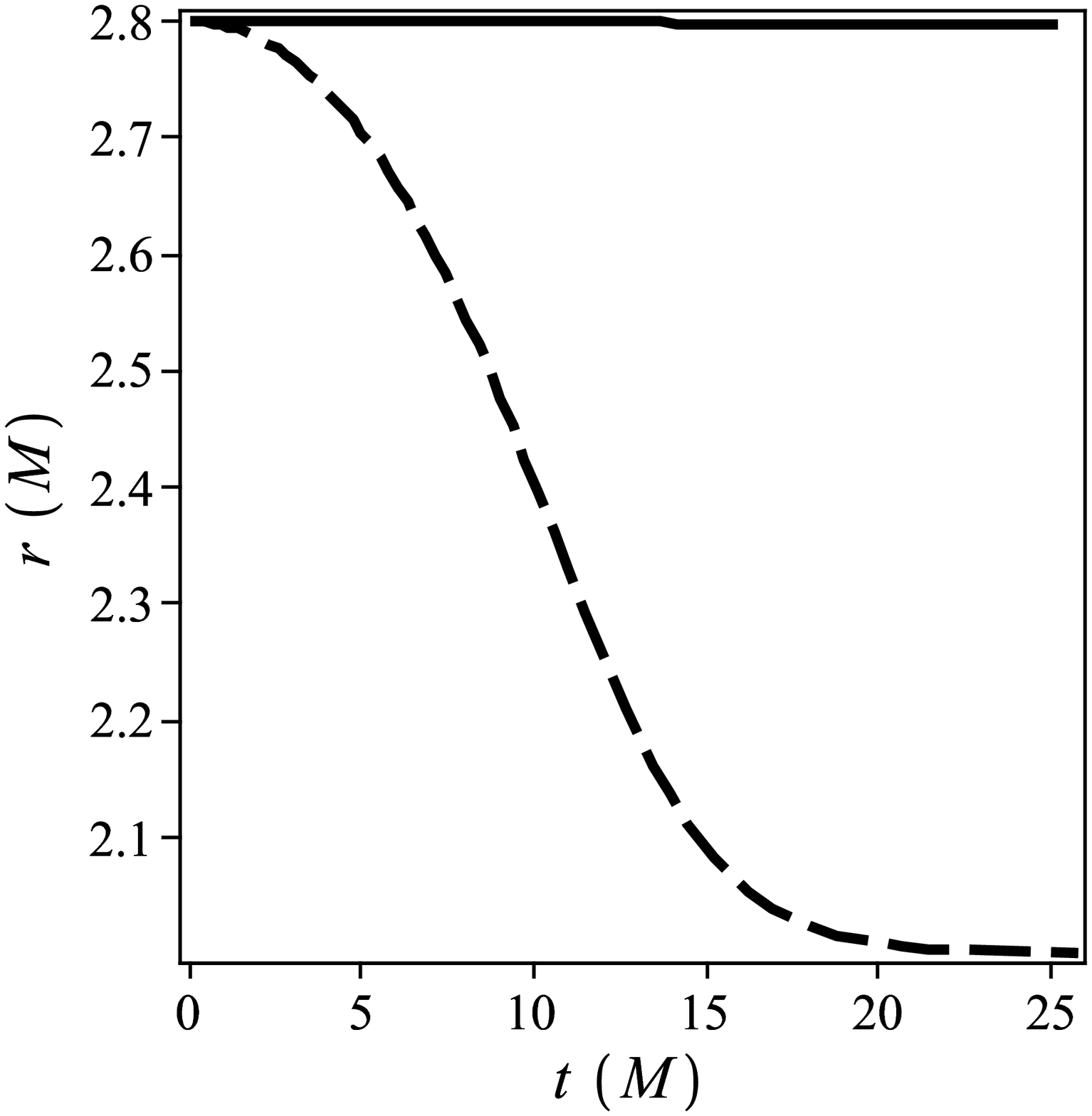}
\caption{\label{1} Radial coordinate vs. coordinate time for the
spinning particle with $S_0/Mm= 10^{-6}$ (solid line) and for the
geodesic motion with the same initial values of the coordinates
and velocity (dashed line).}
\end{figure}

\begin{figure}[!]
\centering
\includegraphics[width=6.5cm]{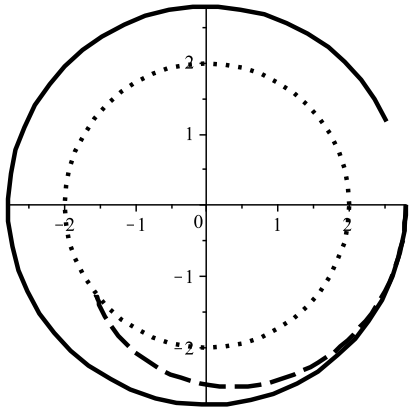}
\caption{\label{1} Trajectories in the polar coordinates of the
spinning particle with $S_0/Mm= 10^{-6}$ (solid line) and spinless
particle (dashed line) with the same initial values of the
coordinates and velocity. The circle with the radius 2 corresponds
to the horizon line.}
\end{figure}

\newpage

\section{Conclusions}

In this paper we obtained the representation of the exact MPD
equations at supplementary condition (3) for the Kerr space-time
by using the constants of the particle's motion, the energy and
angular momentum, together with the differential consequence of
these equations (14). The full set of the corresponding 11
first-order differential equations is presented in (22),
(A.3)--(A.9). The computer integration of these equations is
performed for more simple case of the Schwarzschild space-time.
The possibility using expressions (43), (44) to describe motions
of a spinning particle in this space-time is considered.
Naturally, it is not obvious {\it a priori} that the constants of
motion which are calculated assuming condition (4) can be used for
the correct description of the particle motions by the equations
which were derived using condition (3). According to figures 1--8
it is possible in some approximation. For a more exact description
the expressions for $\hat E$, $\hat J$ from (43), (44) must be
corrected. To get these improved values of $\hat E$ and $\hat J$
one can use of computer search. The suitable values of the pair
$\hat E$, $\hat J$ must give the solutions without large amplitude
oscillations. In some cases instead of (43), (44) it is possible
to use expressions (47), (48) (figures 9--11).

Independent of the results of sections 2, 3 and 5, the important
new information concerning the relationship between the particle's
momentum and velocity according to the MPD equations at condition
(4) is presented in expressions (37)--(40).

In another paper we plan to carry out a more detailed analysis of
the possible choice of values $\hat E$, $\hat J$ and to present
the results of a complex investigation of the highly relativistic
motions of a spinning particle in the Kerr space-time according to
the exact MPD equations (22), (A.3)--(A.9).

\section*{Appendix. Explicit form of the seven equations described in section 3}

First, we stress that the complex explicit form of the seven
equations in a Kerr metric, which were discussed in section 3, is
determined by the much more complicated expressions for the
components of the Riemann tensor and the Christoffel symbols for
this metric than for Schwarzschild's one. We do not write these
expressions here for brevity because they are presented in other
papers, for example, in the Appendix of [8].

We need introduce the notation (in addition to (18)--(21)):
$$
\fl z=y_1^2+\alpha^2\cos^2y_2,\quad q=y_1(y_1-2)+\alpha^2,\quad
 \psi=y_1^2-\alpha^2\cos^2y_2,
$$
$$
\fl \eta=3y_1^2-\alpha^2\cos^2y_2,\quad
 \chi=y_1^2+\alpha^2,\quad \xi=y_1^2-3\alpha^2\cos^2y_2, \eqno
(A.1)
$$
where $\alpha$ is equal to $a/M$. One can see that very quantities
analogous to (A.1) are presented in the expressions for many
components of the Riemann tensor and the Christoffel symbols [8].

To achieve more compact form of the equations we use other
notation as well:
$$
\fl p_1=-zy_5 q^{-1}, \quad p_2=-zy_6,
$$
$$
\fl p_3= [2\alpha
y_1y_8\sin^2{y_2}-y_7(z\chi+2\alpha^2y_1\sin^2{y_2})\sin^2{y_2}
]z^{-1},
$$
$$
\fl p_4=[2\alpha y_1y_7\sin^2{y_2}+y_8(z-2y_1)]z^{-1};
$$
$$
\fl c_1=y_7 y_{10}\sin{y_2}-(z-2y_1)y_6 y_{11}q^{-1}\sin^{-1}y_2,
$$
$$
\fl c_2=y_5 y_{11}(z-2 y_1)q^{-1}\sin^{-1}y_2-y_7 y_9 q\sin y_2,
$$
$$
\fl c_3=(y_6 y_9q-y_5 y_{10})\sin y_2;
$$
$$
\fl d_1=-(2\alpha q^{-1}y_1y_6y_{11}+y_8y_{10})\sin y_2,
$$
$$
\fl d_2=(2\alpha q^{-1}y_1y_5y_{11}+qy_8y_9)\sin y_2,
$$
$$
\fl d_3=(y_5 y_{10}-q y_6 y_9)\sin y_2;
$$
$$
\fl p=2\alpha y_1 y_7 \sin^2{y_2}+(z-2y_1) y_8. \eqno (A.2)
$$
Then the pointed out in section 3 all seven equations can be
written as:
$$
\fl
 y_9\dot y_5+y_{10}\dot y_6+y_{11}\dot y_7=
A-y_9Q_1-y_{10}Q_2-y_{11}Q_3, \eqno (A.3)
$$

$$
\fl
 p_1\dot y_5+p_2\dot y_6+p_3\dot y_7+p_4\dot y_8 =
$$
$$
\fl
 =-p_1Q_1-p_2Q_2-p_3Q_3-p_4Q_4, \eqno (A.4)
$$

$$
\fl c_1\dot y_5+c_2\dot y_6+c_3\dot
y_7=C-c_1Q_1-c_2Q_2-c_3Q_3+\hat E, \eqno (A.5)
$$

$$
\fl d_1\dot y_5+d_2\dot y_6+d_3\dot
y_8=D-d_1Q_1-d_2Q_2-d_3Q_4-\hat J, \eqno (A.6)
$$

$$
\fl  p \dot y_9=\alpha
y_5y_7y_9(\alpha^2-y_1^2)q^{-1}\sin^2{y_2}-\alpha^3 y_1y_6 y_7
y_9z^{-1} \sin^2{y_2} \sin {2y_2}+\alpha^2 y_5 y_8
y_9(y_1-1)q^{-1}
$$
$$
\fl
\times\sin^2{y_2}-0.5\alpha^2y_6y_8y_9(z-2y_1)z^{-1}\sin{2y_2}+
 \alpha y_5^2y_{11}\psi q^{-2}+2 \alpha y_1 y_7^2 y_{11}(y_1 z(z-2
 y_1)
$$
$$
\fl
 -\alpha^2 \psi\sin^2y_2) z^{-2}q^{-1}\sin y_2^2+\alpha y_8^2
y_{11}(z-
 2 y_1)\psi z^{-2}q^{-1} +\alpha y_1y_5 y_7
 y_{10}q^{-1}\sin{2y_2}+0.5\alpha^2y_5
 $$
 $$
\fl
 \times y_8
y_{10}(z- 4 y_1)z^{-1}q^{-1}\sin{2y_2}-2 \alpha y_1y_5 y_6
y_{11}q^{-1}cot y_2
  +2 \alpha y_1^2y_6 y_7 y_{10}z^{-1} \sin^2y_2+
  y_1y_6 y_8
 $$
 $$
\fl
  \times y_{10}(z-2 y_1)z^{-1}+y_7y_8 y_{11}(y_1z(z-2 y_1)^2-\alpha^2\psi(z-
    4 y_1)\sin^2y_2 )z^{-2}q^{-1}+\beta[2\alpha y_5^2y_7(q(z
$$
$$
\fl
    -3y_1^2)+y_1z(y_1-
      1))q^{-2}\sin^2y_2+y_5^2y_8 q^{-2}(3q\psi+\alpha^2z(1-y_1)\sin^2y_2)+
      2\alpha y_1 y_5 y_6 y_7
$$
$$
\fl \times(z+2 \alpha^2\sin^2y_2)q^{-1}\sin{2y_2} +\alpha^2y_5 y_6
y_8(z-4 y_1)q^{-1}\sin{2 y_2}+
 2 \alpha y_1^2y_6^2y_7\sin^2y_2
 $$
 $$
\fl
 +y_1y_6^2y_8(z-2 y_1)-y_8^3(z-2 y_1)\psi z^{-2}+2 \alpha y_7y_8^2\psi(z-3 y_1)z^{-2}\sin^2y_2
  +2 \alpha y_1y_7^3[y_1z\chi(z
  $$
  $$
\fl
 -2 y_1) +\alpha^2\sin^2y_2(zq-2 \alpha^2y_1^2\sin^2y_2+4
y_1^3)]z^{-2}q^{-1}\sin^4y_2+2 \alpha y_1z(y_7\dot y_5 -y_5\dot
y_7)q^{-1}\sin^2y_2
$$
$$
\fl +z(z-
  2 y_1)(y_5\dot y_8-y_8\dot y_5)q^{-1}+y_7^2y_8(y_1z^2(z-
  2 y_1)-\alpha^2\psi(z-6 y_1)\sin^2y_2)z^{-2}\sin^2y_2], \eqno
  (A.7)
$$

$$
\fl p\dot y_{10} = \alpha y_5y_6y_{11}(2y_1^2-z)
q^{-1}-2\alpha^3y_1y_5y_7y_9 z^{-1}\sin^3{y_2}\cos y_2
 +0.5\alpha^2y_5y_8y_9(2 y_1-z)z^{-1}
$$
$$
 \fl
 \times\sin{2 y_2}+
 2\alpha y_1^2 y_5y_7y_{10}z^{-1}\sin^2{y_2}+
   y_1 y_5y_8y_{10}(z-2y_1)z^{-1}-2\alpha y_1y_6^2y_{11}cot y_2
$$
$$
    \fl +\alpha y_1y_7^2y_{11}(z+
    2\alpha^2y_1\sin^2y_2)z^{-2}\sin{2y_2}+2\alpha y_1y_8^2y_{11}(2y_1-z)z^{-2}cot y_2
   +\alpha y_6y_7y_9q
$$
$$
   \fl
   \times(z-4 y_1^2)z^{-1}\sin^2y_2 +
   y_6y_8y_9q(4 y_1^2-z(y_1+1))z^{-1}+
   \alpha y_1y_6y_7y_{10}\sin{2y_2}-
   0.5\alpha^2 y_6y_8y_{10}
$$
$$
   \fl
   \times\sin{2y_2}+y_7y_8y_{11}(z^3+2y_1z(\alpha^2\sin^2y_2-z)-
   8\alpha^2y_1^2\sin^2{y_2})z^{-2}\cot y_2+
    \beta[0.5\alpha^2y_5^2y_8(2y_1
$$
$$
    \fl
    -z)q^{-1}\sin{2y_2}-
     2\alpha^3y_1y_5^2y_7q^{-1}\sin^3y_2\cos y_2 -
     2\alpha y_5y_6y_7\eta\sin^2y_2
      +2 y_5y_6y_8(4 y_1^2-z(y_1+1))
$$
$$
\fl
+\alpha y_1y_6^2y_7(2 z+3\alpha^2\sin^2y_2)\sin{2y_2}
       +\alpha^2y_6^2y_8(z-6 y_1)\sin y_2\cos y_2+2 \alpha y_1z(y_6\dot y_7-y_7\dot
y_6)
$$
$$
\fl
\times\sin^2y_2+z(z-
       2 y_1)(y_6\dot y_8-y_8\dot y_6)
    +\alpha^2y_1y_8^3(z-2 y_1)z^{-2}\sin{2y_2}+
    2 \alpha y_1y_7^3(\chi z^2
$$
$$
\fl +4 \alpha^2y_1z\sin^2{y_2}+
    2\alpha^4y_1\sin^4y_2)z^{-2}\sin^3 y_2\cos y_2+2 \alpha y_1y_7y_8^2(\alpha^2y_1\sin^2y_2-
     \chi(z-2 y_1))z^{-2}
$$
$$
     \fl
     \times\sin{2y_2}+0.5y_7^2y_8(z^3q+
     2 \alpha^2y_1(zq-6 y_1\chi)\sin^2 y_2)z^{-2}\sin{2y_2}], \eqno (A.8)
$$

$$\fl
p\dot y_{11} =-2\alpha^3
y_1y_6y_7y_{11}z^{-1}\sin^3{y_2}\cos{y_2}+\alpha y_5y_7y_{11}
 [(4y_1^2-z)\chi-2\alpha^2 y_1^2\sin^2y_2-4y_1^3 ]z^{-1}q^{-1}
$$
$$
\fl
\times\sin^2y_2
 +y_5y_8y_{11}[y_1(z-2 y_1)^2+\alpha^2(z-2y_1^2)\sin^2{y_2}]z^{-1}q^{-1}
 +y_6y_8y_{11}[z(z-2 y_1)+2\alpha^2 y_1
 $$
$$ \fl
\times\sin^2{y_2}]z^{-1}\cot y_2-\alpha
y_7^2y_9q(\alpha^2\psi+y_1^2(z+2 y_1^2))z^{-2}\sin^4y_2
 +y_7y_8y_9q(zy_1(2 y_1-z)+\psi(\chi
 $$
 $$
 \fl
 +\alpha^2\sin^2{y_2}))z^{-2}\sin^2y_2-\alpha y_8^2 y_9q\psi
z^{-2}\sin^2y_2
  +2 \alpha^3y_1y_7^2y_{10}q z^{-2}\sin^5{y_2}\cos y_2-0.5y_7y_8y_{10}q
  $$
  $$
  \fl
  \times(z^2+4 \alpha^2y_1\sin^2
  {y_2})z^{-2}\sin{2y_2}
  +\alpha y_1y_8^2y_{10}qz^{-2}\sin{2y_2}+\beta[-2\alpha y_5y_7^2
  (\chi\psi+2y_1^2z)z^{-1}\sin^4{y_2}
  $$
  $$
  \fl
  +4\alpha^3y_1y_6y_7^2qz^{-1}\sin^5y_2\cos y_2-2 \alpha y_5y_8^2\psi z^{-1}\sin^2y_2
  +2\alpha y_1y_6y_8^2q z^{-1}\sin{2y_2}
   +zq(y_7\dot y_8
  $$
$$
\fl -y_8\dot y_7)\sin^2{y_2}-y_6 y_7y_8zq\sin{2y_2}
   -2y_5y_7y_8(y_1z(z-
   2y_1)-\psi(\chi+\alpha^2\sin^2{y_2}))z^{-1}\sin^2y_2], \eqno
   (A.9)
$$
where
$$
 \fl
 A=-2\alpha y_5y_6y_9y_{10}\eta z^{-3} \cos y_2
 -2\alpha y_5y_7y_9y_{11}\eta z^{-4}(\chi+2\alpha^2\sin^2{y_2})\cos
y_2
$$
$$ \fl
-6\alpha y_1y_6y_7y_9y_{11}q\xi z^{-4}\sin y_2
+6y_1y_6y_8y_9y_{11}q\xi z^{-4}\sin^{-1}{y_2} + 6\alpha
y_1y_7^2y_9y_{10}q\xi\chi z^{-5}\sin^3{y_2}
$$
$$
\fl -6y_1y_7y_8y_9y_{10}q\xi
z^{-5}(\chi+\alpha^2\sin^2{y_2})\sin{y_2} +6\alpha
y_1y_8^2y_9y_{10}q\xi z^{-5}\sin{y_2}
$$
$$
\fl +2\alpha y_6y_7y_{10}y_{11}\eta
z^{-4}(2\chi+\alpha^2\sin^2{y_2})\cos y_2 -6\alpha
y_1y_5y_7y_{10}y_{11}\xi\chi q^{-1}z^{-4}\sin y_2
$$
$$
\fl +6\alpha^2 y_1y_5y_8y_{10}y_{11}\xi q^{-1}z^{-4}\sin y_2 +
\alpha y_6^2y_9^2q\eta z^{-3}\cos y_2
$$
$$
\fl +\alpha y_7^2y_9^2\eta
q(\chi^2+2q\alpha^2\sin^2{y_2})z^{-5}\sin^2{y_2}\cos y_2
$$
$$
\fl -2\alpha^2 y_7y_8y_9^2\eta q(3\chi-4y_1)z^{-5}\sin^2{y_2}\cos
y_2+ \alpha\eta q(2q+\alpha^2\sin^2{y_2})z^{-5} y_8^2y_9^2 \cos
y_2
$$
$$
\fl+\alpha\eta y_5^2y_{10}^2 q^{-1}z^{-3}\cos y_2
-\alpha\eta(2\chi^2+ \alpha^2
q\sin^2{y_2})y_7^2y_{10}^2z^{-5}\sin^2{y_2}\cos
y_2
$$
$$
\fl -\alpha\eta(q+2\alpha^2\sin^2{y_2}) y_8^2y_{10}^2 z^{-5}\cos
y_2+ 2\alpha^2\eta (3\chi-2y_1)\times
y_7y_8y_{10}^2z^{-5}\sin^2{y_2}\cos y_2
$$
$$
\fl+\alpha\eta(q+2\alpha^2\sin^2{y_2})y_5^2y_{11}^2
q^{-2}z^{-3}\sin^{-2}{y_2}\cos y_2
-\alpha\eta(2q+\alpha^2\sin^2{y_2})
$$
$$
\fl \times y_6^2y_{11}^2 q^{-1}z^{-3}\sin^{-2}{y_2}\cos y_2
+6\alpha\xi y_1y_5y_6y_{11}^2 q^{-1}z^{-3}\sin^{-1}{y_2}
$$
$$
\fl -4\alpha^3y_1^2\eta y_7^2y_{11}^2 q^{-1}z^{-5} \sin^2{y_2}\cos
y_2
$$
$$
\fl -2\alpha^2y_1\eta(q-\alpha^2\sin^2{y_2})(1+
\sin^2{y_2})y_7y_8y_{11}^2 q^{-1}z^{-5}\cos y_2
$$
$$
\fl -\alpha\eta(z-2y_1)(q-\alpha^2\sin^2{y_2})y_8^2y_{11}^2
q^{-1}z^{-5} \cos y_2; \eqno (A.10)
$$

$$
\fl Q_1=(y_1 q-z(y_1-1)) y_5^2 z^{-1}q^{-1}
  -qy_1y_6^2 z^{-1}-q(y_1z^2-\alpha^2\psi\sin^2{y_2})y_7^2
  z^{-3}\sin^2{y_2}
$$
$$
+q\psi y_8^2 z^{-3}-\alpha^2y_5 y_6 z^{-1}\sin 2y_2-
  2\alpha q\psi y_7y_8\ z^{-3}\sin^2{y_2},
$$
$$
\fl
Q_2=-0.5\alpha^2y_6^2 z^{-1}\sin{2y_2}+0.5\alpha^2y_5^2
z^{-1}q^{-1}\sin{2y_2}
  -0.5y_7^2(z^2\chi
$$
$$
+2\alpha^2y_1(\chi+z)\sin^2{y_2})z^{-3}\sin{2y_2}-
  \alpha^2y_1y_8^2 z^{-3}\sin{2y_2}
  +2y_1y_5 y_6z^{-1} +2\alpha y_1 y_7 y_8\chi z^{-3}\sin{2y_2},
$$
$$
\fl Q_3=2y_5y_7(y_1z(z-2y_1)-\alpha^2\psi\sin^2{y_2})z^{-2}q^{-1}+
2\alpha y_5y_8\psi z^{-2}q^{-1}
$$
$$
\fl +2y_6y_7(z^2+2\alpha^2 y_1\sin^2{y_2})z^{-2}\cot y_2 - 4\alpha
y_1y_6y_8z^{-2}\cot y_2.
$$
$$
\fl Q_4=-2\alpha y_5
y_7(2y_1^2z+\psi\chi)z^{-2}q^{-1}\sin^2{y_2}+2y_5y_8\psi\chi
z^{-2}q^{-1}
$$
$$
\fl
 +2\alpha^3y_1y_6y_7 z^{-2}\sin^2{y_2}\sin{2y_2}-
2\alpha^2y_1y_6y_8 z^{-2}\sin{2y_2}; \eqno (A.11)
$$

$$
 \fl
 C=-(1-2y_1z^{-1})y_8-2\alpha y_1 y_7 z^{-1}\sin^2{y_2}
 $$
 $$
  \fl
  +2\alpha^2 qy_1y_7y_9 z^{-3}\sin^2{y_2}\cos{y_2}-
2\alpha qy_1y_8y_9 z^{-3}\cos{y_2}
$$
$$
\fl
+\chi\psi y_7y_{10}z^{-3}\sin{y_2} -\alpha y_8y_{10}\psi
z^{-3}\sin{y_2}
$$
$$
\fl -2\alpha^2y_1y_5y_{11}q^{-1}z^{-2}\cos{y_2}- y_6y_{11}\psi
z^{-2}\sin^{-1}{y_2}; \eqno (A.12)
$$

$$
\fl D=-2\alpha^3 y_1 y_7 y_9 z^{-3}\sin^4{y_2}\cos y_2 +
q(z^2+2\alpha^2 y_1\sin^2{y_2})
$$
$$ \fl
\times y_8 y_9 z^{-3}\cos y_2 -\alpha y_7 y_{10}(\chi\psi+2y_1^2
z)z^{-3}\sin^3{y_2}
$$
$$
\fl -y_8 y_{10}(y_1z(z-2y_1)-\alpha^2\psi\sin^2{y_2})z^{-3}\sin
y_2
$$
$$
\fl+2\alpha y_1 y_5 y_{11} \chi q^{-1}z^{-2}\cos y_2 + \alpha\psi
y_6 y_{11} z^{-2}\sin y_2
$$
$$
\fl+p^{-1}[2\alpha y_1 y_7^2(z\chi+2\alpha^2
y_1\sin^2{y_2})z^{-1}\sin^2{y_2} + qzy_7 y_8
$$
$$
\fl \times(1-8\alpha^2 y_1^2 q^{-1} z^{-2}\sin^2{y_2})-2\alpha
y_1(z-2y_1)z^{-1} y_8^2]\sin^2{y_2}; \eqno (A.13)
$$

$$
\fl \beta=y_5y_9+y_6y_{10}+y_7y_{11}. \eqno (A.14)
$$

We stress that both equations (A.3)--(A.9) and expressions
(A.10)--(A.13) become much simpler in the case of the
Schwarzschild space-time, when $\alpha\equiv a/M=0$. Then, for
example, instead of long expression (A.10) we have

$$
 \fl
 A=6y_1y_6y_8y_9y_{11}q\xi z^{-4}\sin^{-1}{y_2}
-6y_1y_7y_8y_9y_{10}q\xi
z^{-5}(\chi+\alpha^2\sin^2{y_2})\sin{y_2}. \eqno (A.15)
$$


\section*{References}


\begin{thebibliography}{100}
\bibitem{1} Fock V and Ivanenko D 1929 \textit{Z. Phys.}  \textbf{54} 798

Fock V 1929 \textit{Z. Phys.}  \textbf{57} 261

Weyl H 1929 \textit{Proc. Nat. Acad. Sci. USA}  \textbf{15} 323


\bibitem{2} Mathisson M 1937
\textit{Acta Phys. Pol.} \textbf{6} 163

\bibitem{3}
Wong S 1972 \textit{Int. J. Theor. Phys.}  \textbf{5} 221

Kannenberg L 1977 \textit{Ann. Phys.(N.Y.)} \textbf{103} 64

Catenacci R and Martellini M 1977 \textit{Lett. Nuovo Cimento}
\textbf{20} 282

Audretsch J 1981 \textit{J. Phys.} A  \textbf{14} 411

Gorbatsievich A 1986 \textit{Acta Phys. Pol.} B  \textbf{17} 111

Barut A and Pavsic M 1987 \textit{Class. Quantum Grav.} \textbf{4}
41

Cianfrani F and Montani G 2008 \textit{Europhys. Lett.}
\textbf{84} 30008

Cianfrani F and Montani G 2008 \textit{Int. J. Mod. Phys.} A
\textbf{23} 1274

Obukhov Yu Silenko A and Teryaev O 2009 \textit{Phys. Rev.} D
\textbf{80} 064044

\bibitem{4} Papapetrou A 1951
\textit{Proc. R. Soc.} A \textbf{209} 248

\bibitem{5} Dixon W G
1970 \textit{Proc. R. Soc.} A \textbf{314} 499

Dixon W G 1973 \textit{Gen. Rel. Grav.}  \textbf{4} 199

 Dixon W G 1974
\textit{Philos. Trans. R. Soc.} A \textbf{277} 59

Dixon W G 2008 \textit{Acta Phys. Pol. B. Proc. Suppl.} \textbf{1}
27

\bibitem{6} Steinhoff J and Puetzfeld D 2010 \textit{Phys. Rev.} D \textbf{81}
044019

\bibitem{7} Corinaldesi E and Papapetrou A 1951
\textit{Proc. R. Soc.} A \textbf{209} 248

\bibitem{8} Semerak O
1999 \textit{Mon. Not. R. Astron. Soc.}  \textbf{308} 863

\bibitem{9} Kyrian K and Semerak O
2007 \textit{Mon. Not. R. Astron. Soc.}  \textbf{382} 1922

\bibitem{10} Schiff L
1960 \textit{Phys. Rev. Lett.}  \textbf{4} 219

\bibitem{11} Wald R 1972 \textit{Phys. Rev.} D \textbf{6} 406

\bibitem{12} Plyatsko R and Vynar A 1982 \textit{Sov. Phys.-Dokl.} \textbf{27}
328

\bibitem{13} Plyatsko R 1998 \textit{Phys. Rev.} D \textbf{58}
084031

Plyatsko R and Bilaniuk O 2001 \textit{Class. Quantum Grav.}
\textbf{18} 5187

Plyatsko R  2005 \textit{Class. Quantum Grav.} \textbf{22} 1545

\bibitem{14} Plyatsko R, Stefanyshyn O and Fenyk M
2010 \textit{Phys. Rev.} D \textbf{82} 044015

\bibitem{15} Stuchl{\'i}k Z 1999 \textit{Acta Phys. Slovaca}  \textbf{49}
319

Stuchl{\'i}k Z and Hled{\'i}k S 2001 \textit{Phys. Rev.} D
\textbf{64} 104016

Stuchl{\'i}k Z and Kov\'a\v{r} J 2006 \textit{Class. Quantum
Grav.} \textbf{23} 3935

Mortazavimanesh M and Mohseni M 2009 \textit{Gen. Rel. Grav.}
\textbf{41} 2697

\bibitem{16} Bi\v{c}{\'a}k J Stuchl{\'i}k Z and Balek V 1989 \textit{Bull. Astronom.
Inst. Czechoslovakia}  \textbf{40} 65

Bonnor W B 1993 \textit{Class. Quantum Grav.} \textbf{10} 2077

Stuchl{\'i}k Z Bi\v{c}{\'a}k J and Balek V 1999 \textit{Gen. Rel.
Grav.} \textbf{31} 53

\bibitem{17} Pirani F A E
1956 \textit{Acta Phys. Pol.} \textbf{15} 389

\bibitem{18} Tulczyjew W
1959 \textit{Acta Phys. Pol.} \textbf{18} 393

\bibitem{19} Barker B M and O'Connel R F 1979
\textit{Gen. Rel. Grav.}  \textbf{11} 149

\bibitem{20} Aleksandrov A N
1991 \textit{Kinem. Fiz. Nebesn. Tel} \textbf{7} 13

\bibitem{21} Mashhoon B 1975 \textit{Ann. Phys.}  \textbf{89} 254

Bini D Cherubini C Geralico A and Jantzen R T 2006 \textit{Int. J.
Mod. Phys.} D \textbf{15} 737

\bibitem{22} Tod K P and de Felice F
1976 \textit{IL Nuovo Cimento} \textbf{34} 365

\bibitem{23} Hojman R and Hojman S 1977
\textit{Phys. Rev.} D \textbf{15} 2724

\bibitem{24} Suzuki S and Maeda K 1998
\textit{Phys. Rev.} D \textbf{58} 023005

\bibitem{25} Hartl M 2003
\textit{Phys. Rev.} D \textbf{67} 024005; 104023

\bibitem{26} Landau L D and  Lifshitz E M  1971 \textit{The Classical Theory of
Fields} (Massachusetts: Addison-Wesley)

\bibitem{27} Chandrasekhar S   1983 \textit{The Mathematical Theory of Black Holes}
(Oxford: Oxford University Press)

\bibitem{28} Mathisson M 1937
\textit{Acta Phys. Pol.} \textbf{6} 218

\bibitem{29} Weyssenhoff J and Raabe A 1947 \textit{Acta. Phys. Pol.}
\textbf{9} 7

\bibitem{30} M\"oller C 1949 \textit{Commun. Dublin Inst.
Advan. Studies} A  \textbf{5} 3

\bibitem{31} Taub H 1964 \textit{J. Math.
Phys.}   \textbf{5} 112

\bibitem{32} Mashhoon B 1971 \textit{J. Math. Phys.} \textbf{12}
1075

\bibitem{33} Ragusa S and Bailyn M 1995 \textit{Gen.
Rel. Grav.}  \textbf{27} 163

\bibitem{34} Chicone C,  Mashhoon B and Punsly B 2005
\textit{Phys. Lett.} A \textbf{343} 1

\bibitem{35} Mashhoon B and Singh D 2006
\textit{Phys. Rev.} D \textbf{74} 124006

\bibitem{36} Singh D 2008
\textit{Phys. Rev.} D \textbf{78} 104028

\bibitem{37} Han V-B 2008  \textit{Gen.
Rel. Grav.}  \textbf{40} 1831

\bibitem{38} Obukhov Yu and Puetzfeld D 2011
\textit{Phys. Rev.} D \textbf{83} 044024

\bibitem{39} M\"oller C 1972 \textit{The Theory of
Relativity} (Oxford: Oxford University Press)

\bibitem{40} Plyatsko R and Stefanyshyn O 2008
\textit{Acta Phys. Pol.} B \textbf{39} 23

\bibitem{41} Abramowicz M A and  Calvani M 1979 \textit{Mon. Not. R. Astron. Soc.}
\textbf {189}, 621

 Svirskas K, Pyragas K and Lozdiene A 1988
\textit{Astrophys. Space Sci.} \textbf {149}, 39

 Rietdijk R H
and van Holten  J W 1993  \textit{Class. Quantum Grav.} \textbf
{10}  575

 Hossain Ali M and Mainuddin A 2000 \textit{Ann. Phys.}
\textbf {282} 157

 Bini D,  de Felice F de and  Geralico A
2004 \textit{Class. Quantum Grav.} \textbf {21} 5441

 Faruque
S B 2004 \textit{Phys. Lett.}  \textbf {327} 95

 Burko L M 2004
\textit{Phys. Rev.} D \textbf {69} 044011

 Bini D, Geralico A and
de Felice F  2005  \textit{Int. J. Mod. Phys.} D  \textbf {14}
1793

 Bini D, de Felice F, Geralico A and Jantzen R T 2005 \textit{Class.
Quantum Grav.} \textbf{22} 2947

Wang J and Y.-J. Wang Y-J 2005 \textit{Chin. Phys. Lett.} \textbf
{22} 539

 Bini D, de Felice F, Geralico A and Jantzen R T 2006
\textit{Class. Quantum Grav.} \textbf {23} 3287

 Mohseni M 2010
\textit{Gen. Rel. Grav.} \textbf {42} 2477



\end{thebibliography}
\end{document}